# Cymatics Cup: Shape-Changing Drinks by Leveraging Cymatics

Weijen Chen
Keio University, Graduate School of Media Design
Yokohama, Kanagawa, Japan
weijen@kmd.keio.ac.jp

Yang Yang
Keio University, Graduate School of Media Design
Yokohama, Japan
seanlin@kmd.keio.ac.jp

Kao-Hua Liu
Tokyo University, Research Center for Advanced Science and Technology
Tokyo, Japan
maarkliu@star.rcast.u-tokyo.ac.jp

Yun Suen Pai
University of Auckland
Auckland, New Zealand
yspai1412@gmail.com

Junichi Yamaoka
Keio University, Graduate School of Media Design
Yokohama, Japan
yamaoka@kmd.keio.ac.jp

Kouta Minamizawa
Keio University, Graduate School of Media Design
Yokohama, Japan
kouta@kmd.keio.ac.jp

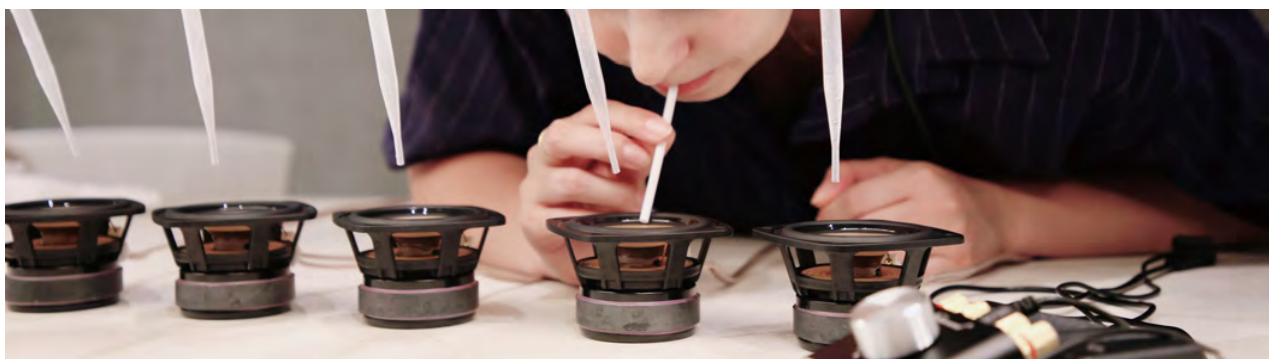

(a)

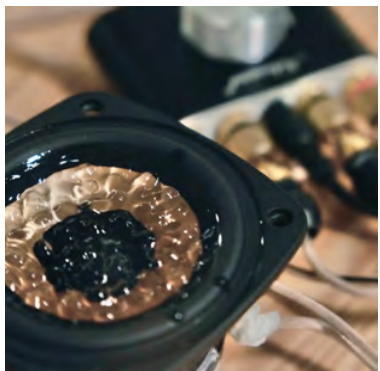

(b)

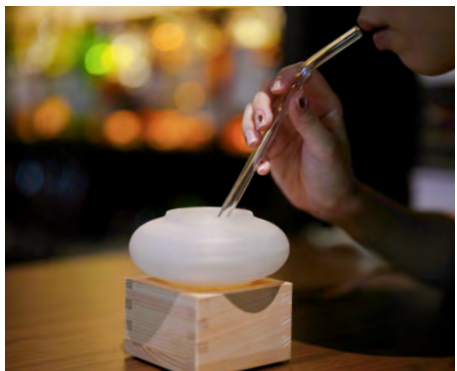

(c)

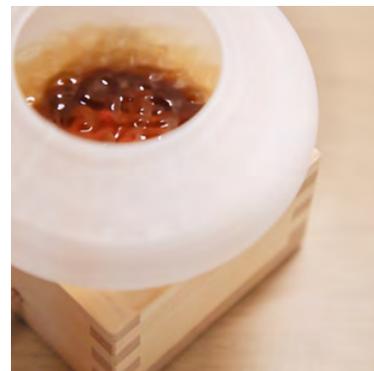

(d)

Figure 1: (a) Cymatics Seasoning Experiment (b) Cymatics Phenomenon (c) Cymatics Cup (d) The Beverage in Cymatics Cup

## ABSTRACT

To enhance the dining experience, prior studies in Human-Computer Interaction (HCI) and gastrophysics have demonstrated that modifying the static shape of solid foods can amplify taste perception. However, the exploration of dynamic shape-changing mechanisms in liquid foods remains largely untapped. In the present study, we employ cymatics, a scientific discipline focused on utilizing sound frequencies to generate patterns in liquids and particles—to augment the drinking experience. Utilizing speakers, we dynamically reshaped liquids exhibiting five distinct taste profiles and evaluated resultant changes in taste perception and drinking experience. Our research objectives extend beyond merely augmenting taste from visual to tactile sensations; we also prioritize the experiential aspects of drinking. Through a series of experiments and workshops, we revealed a significant impact on taste perception and overall drinking experience when mediated by cymatics effects. Building upon these findings, we designed and developed tableware to integrate cymatics principles into gastronomic experiences.

## KEYWORDS

food, cymatics, crossmodal correspondences, gastrophysics

**ACM Reference Format:**
Weijen Chen, Yang Yang, Kao-Hua Liu, Yun Suen Pai, Junichi Yamaoka, and Kouta Minamizawa. 2024. Cymatics Cup: Shape-Changing Drinks by Leveraging Cymatics. In *Proceedings of the CHI Conference on Human Factors in Computing Systems (CHI '24), May 11–16, 2024, Honolulu, HI, USA*. ACM, New York, NY, USA, 19 pages. https://doi.org/10.1145/3613904.3642920

## 1 INTRODUCTION

Dining is one of the most fundamental aspects of our daily life. While it was once solely for survival, people nowadays seek a more enjoyable and fulfilling dietary lifestyle. In 2017, at the SKS Conference[1] held in the United States, Rebecca Chesney[2] from the "Institute for the Future," a think tank with over fifty years of experience in future studies, presented a discourse proposing that modern food embodies twelve new values (or life demands concerning diet), including "Discovery," "Experimentation," "Novelty," "Individuality," and "Inclusivity," which have already surpassed traditional notions of "Efficiency" and "Convenience" [111].

In the field of gastrophysics, research aimed at enhancing the dining experience aligns seamlessly with Chesney's proposition of modern food's newfound value. Gastrophysics can be defined as the scientific study of those factors that influence our multisensory experience while tasting food and drink. It combines the science of gastronomy and psychophysics, and it has been found that various sensory factors, including smell, sight, sound, touch, and the overall atmosphere, influence human taste perception and dining experience [80]. Visual perception is one of the commonly explored aspects in influencing taste sensation. While we have learned from Shepherd [77] that olfaction is a key factor in affecting taste, Wang et al. [100] have also discovered a strong correlation between Steinway pianos and taste. Additionally, factors such as temperature, vibration [49, 90], electric currents[3] [53, 60, 93], and weight sensation [26] can influence the dining experience. Despite these findings, we have identified further potential for exploring the impact of visual perception on the dining experience. As a result, within the field of HCI and Human-Food Interaction (HFI), numerous studies have emerged that leverage Augmented Reality (AR) to enhance the dining experience, capitalizing on the impact of visual perception [61].

In terms of the visual aspect, we are aware that the shape of food can strongly influence taste perception [80]. Deroy and Valentin [15] have highlighted that individuals frequently tend to establish an association between bitter foods and beverages and thin or angular shapes, while sweetness and voluminousness are commonly linked with more rounded forms. This understanding has led to the exploration of 3D printing food to alter the eating experience through shape-changing [45]. There are also methods such as projection mapping that can dynamically change the appearance of food by layering images [86], enhancing the dining experience. Nevertheless, there are limitations when it comes to altering the shape of liquid compared to solid food. While previous research has conducted crossmodal experiments on liquid drinking experiences, such as changing beverage colour [80], there has been limited focus on altering the shape of liquid beverages, especially for dynamical shape-changing. Additionally, the augmented eating approach raises concerns about the inconvenience of utilizing wearable devices in daily life and the safety issues for children when exposed to electric currents[4]. Consequently, the initial research questions are as follows: How can we change the shape of liquid food, such as beverages or alcohol, and explore if augmented eating also works with shape-changing for liquid food? Is there a way to design a convenient drinking tool without the risk of electric current for younger people? Cymatics may offer a solution.

Cymatics is a science that utilizes sound frequencies and vibrations to influence particles, causing them to exhibit visible patterns and shapes on a material [63]. In this study, we propose a new approach to drinking using cymatics. The Cymatics Cup enables the shape of liquid to be changed without the need for any extra wearable devices while drinking. Participants can conveniently drink from the Cymatics Cup without any risk of electric current, similar to how they would in a café or bar, thus not requiring any significant changes to their drinking behaviour.

Originally, we intended to investigate the visual impact of cymatics on taste perception. However, we later discovered that cymatics encompasses three variables: visual, haptic, and auditory, as it is the study of making sound and vibration visible [1]. To gain a better understanding of how cymatics influences taste perception, we plan to explore these different variables. Nevertheless, since Wang et al. [100] have already conducted experiments on the effects of sound pitch on taste perception, our study will focus on the influence of visual and tactile aspects on the drinking experience. This includes aspects such as taste perception and the sensations experienced while drinking. Inspired by previous studies on the visual impact on food, our hypothesis is that the shapes generated by cymatics will yield similar results as seen in previous studies. For instance, round-shaped food is perceived as sweeter [15], and similarly, big round chunk shapes generated by cymatics will also be perceived as sweeter. Other research shows that vibrations can enhance wine tasting [49], and we hypothesize that cymatics will yield similar results when it comes to enhancing the drinking experience. After conducting several study works and pilot experiments on drinking with cymatics, we have specifically hypothesized the following:

(1) H1: The visual impact generated by cymatics can enhance the taste perception of beverages.
(2) H2: The tactile impact generated by cymatics can enhance the taste perception of beverages.
(3) H3: Cymatics affects the texture, appetite, and enjoyment of the drinking experience.

---

[1]https://www.smartkitchensummit.com
[2]https://www.smartkitchensummit.com/rebeccachesney
[3]https://www.kirinholdings.com/en/newsroom/release/2022/0411_01.html

[4]https://gourmet.watch.impress.co.jp/docs/news/1438322.html

To test our assumptions, we divided the testing into two phases. First, in our user study 1, we used speakers as drinking containers to test the influence of different cymatics frequencies on taste perception while drinking various solutions. In the second phase, user study 2 focusing on the feeling of drinking through real beverage testing to observe its effects in a real scenario.

Prior gastrophysics, neurogastronomy, and HFI studies have shown that multi-sensory aspects affect humans' dining experience. Cymatics opens a new dimension in the study of cross-modal perception, exploring new possibilities to enhance our drinking rituals. Overall, our research goal is to provide a new perspective on the multi-sensory aspects of taste and enhance the drinking experience. In the future, this discovery may be applied to nutritional liquid food consumption, improving the appetite of individuals who require specific nutrition intake. Our four main contributions are as follows:

(1) We integrate the research fields of gastrophysics, neurogastronomy, and HCI research, especially in the utilization of cymatics, which provides a novel connection between cymatics and the dining experience.
(2) We not only focus on taste perception in cymatics but also consider the joyfulness that people experience during cymatics drinking.
(3) We designed a prototype of the Cymatics Cup to optimize the drinking experience with the speaker, laying the groundwork for establishing a new drinking ritual in the future.
(4) We composed five demo frequencies for seasoning, which we refer to as 'Cymatics Seasoning.' These frequencies allow people to experience various tasting effects and enhance their drinking experience.

There have been very few prior cymatics toolkits developed to enable users to experience the effects of cymatics on their drinking. We hope that these demos of the Cymatics Cup and Cymatics Seasoning can serve as the first steps towards creating a cymatics toolkit that offers a new perspective on drinking. Our research goal is to utilize this toolkit not only to enhance taste perception but also to stimulate aspects of the drinking experience, such as happiness, desire, and mouthfeel, which were initially of concern to us. The toolkit will undergo further development for practical applications in the future. In this study, we employed standard speakers in user study 1 and user study 2 to test our hypotheses, and the next phase of our research involves applying the toolkit in real-world scenarios.

## 2 RELATED WORKS

In this section, we conduct a comprehensive review of previous research, focusing on two main areas: 1) gastrophysics and neurogastronomy focuses on investigating the effects of visual, auditory, and tactile stimuli, and 2) the realm of HCI encompassing food-related tools, often referred to as Human-Food Interactions. Our work presents a distinctive approach by synergistically integrating the principles of frequency manipulation, known as cymatics, with established HCI community research on food-related aspects. This amalgamation introduces a novel avenue within the domain of food augmentation.

### 2.1 Visual, Auditory, and Tactile Stimuli on Food

In order to understand the underlying principles of HFI, it is essential to examine gastrophysics and neurogastronomy research, which investigate the impact of various sensory stimuli on taste perception. However, because we are specifically interested in the crossmodal effects of cymatics on taste perception and dining experience, among the numerous sensory stimuli, we will focus on gaining a deeper understanding of visual, auditory, and tactile stimuli. Numerous studies in this domain have underscored the pivotal role of colour in taste perception. A seminal experiment in 2001, conducted by Morrot et al. [57] examined red and white wines, serving as an illustrative example of colour's impact on odour and flavour perception. Their work demonstrated that the addition of red colouring to white wine led tasters to perceive it as red wine erroneously. Johnson et al. [32] found that in darker red-coloured solutions, the perceived sweetness was 2–10% greater than in lighter reference solutions, even when the actual sucrose concentration was 1% less. Spence et al. [83] conducted experiments on colour pairing with five taste modalities: sweet, sour, bitter, salty, and umami, confirming the correlation between colour and taste perception. The shape of food is another crucial aspect in the realm of visual influence. As evidenced by Cadbury's decision in 2013 to alter the shape of their iconic Dairy Milk chocolate bar, even slight changes in shape can significantly alter taste perception [20]. Additionally, Nendo, a renowned Japanese design company, showcased the "chocolatexture[5]" collection at Maison & Objet Paris[6] in 2015, which featured chocolates of varying shapes to enhance the eater's sensory experience. Furthermore, G. Van Doorn et al. [95] found that angular shapes had a greater influence on people's expectations of drink likability, bitterness, and quality compared to rounded shapes, as observed in caffè latte toppings.

Due to the discussion of the impact of shape on taste perception, numerous studies have explored crossmodal correspondences, particularly those between sweetness and roundness, as well as sourness, bitterness, and angularity. Montejo et al. [75] extended this research by discovering that shapes characterized by angularity, asymmetry, and a greater number of elements were more likely to be judged as both unpleasant and sour. It was further confirmed by Turoman et al. [92] that shapes with the highest symmetry were perceived as the sweetest, least bitter, most pleasant, and least threatening. Additionally, Harrar and Spence [22] found in their experiments that food was rated as the saltiest when sampled from a knife rather than from a spoon, fork, or toothpick. Extending the exploration of the association between shape and diet, Velasco et al. [97] discovered a close relationship between words, shapes, and tastes, such as round shapes with the taste word "sweet" and angular shapes with the taste words "salty," "sour," and "bitter." Summarizing the aforementioned studies on the visual and shape-related influences on taste perception, it becomes evident that the form of food significantly impacts the dining experience. However, in experiments involving changes in shape and contour, most studies have utilized solid foods for testing. Although some scholars have explored alterations in the contour of liquids through changes in the shape of drinking vessels and observed increased consumption[4, 42, 82], we note that there are still limitations in altering the shape and contour of the liquid itself without relying on modifications to the shape of the containing vessel.

---

[5] https://www.nendo.jp/en/works/chocolatexture-2/
[6] https://www.maison-objet.com/en/paris

Beyond the visual domain, previous research has highlighted the auditory aspect's potential to augment the culinary experience. For instance, the perception of food crispness is primarily informed by the sound generated when biting and breaking down food (e.g., potato chips) with one's teeth. This auditory dimension contributes to the perceived palatability of foods like potato chips, an experience that may be attributed to the neocortical hierarchy [77]. Notably, Knöferle's study [39] manipulated the sounds produced by a Nespresso coffee machine, influencing participants' descriptions of the resulting coffee. Adjusting the machine's noise levels led to significant changes in perceived taste. In addition to research on machine noise, studies have also been conducted on airline cabin noise. Yan et al. [104] found that the perceived intensity of sweet taste decreased progressively, while the perception of umami taste increased during the experimental sound condition, and this effect became more pronounced with higher concentrations. In the realm of human voice research, Kawai et al. [36] also discovered that human voice plays a crucial role in the perceived taste of food and consumption. Music also plays a role; specific musical pieces have been shown to accentuate the flavours of certain foods and beverages [24, 50, 73, 84, 99]. For example, a specially designed bitter soundtrack was rated as tasting significantly more bitter than when exactly the same foodstuff was consumed with sweet soundtrack [14]. Guetta and Loui [21] also designed an experiment involving four different flavours of custom-made chocolate ganache, demonstrating that participants can match music clips with the corresponding taste stimuli with above-chance accuracy. A previous study unveiled a nuanced relationship between pitch and taste perception, where increased taste intensity was associated with higher or lower pitches [100]. Furthermore, it is noteworthy that Simner et al. [78] found that the sweet taste was judged to match smoother, more continuous vowel sounds than the bitter and sour tastes, which were judged to match more staccato sounds.

Equally important, as Verhagen and Engelen outlined in 2016, tactile sensations contribute to the tasting experience. Temperature variations can significantly affect the perceived texture of food in the mouth, potentially arising from both peripheral and central mechanisms. Lower temperatures can mitigate the burning sensation caused by compounds like capsaicin, whereas higher temperatures can enhance it. Moreover, capsaicin's influence extends to altering the perception of heat and cold. The interplay between touch and taste reveals that sweetness imparts a stickier sensation to food, while acidity elicits the opposite effect [77]. Additionally, unlike the tactile sensations within the oral cavity, external sensory perceptions from various sources during dining can also influence our dining experiences. Through crossmodal research, Piqueras-Fiszman and Spence [66] have found that using a heavier spoon, as opposed to a lighter one, leads to higher ratings in the tasting of yogurt. Texture also holds sway over the eating experience, as indicated by Biggs et al.'s findings [7] that the texture of the serving plate can influence perceived spiciness. In addition to approaching from the perspective of neurogastronomy and crossmodal effects, Chung et al. [13] found that the effect of vibration on some physico-chemical characteristics of a commercial red wine led to a slight increase in acidity. Meanwhile, Salek et al. [74] concluded that mechanical vibration can convert mechanical energy into thermal energy and enhance the hardness, storage modulus, and viscosity of cheese sauce, suggesting a potential indirect impact on the sensory properties of cheese sauce. Regarding the aforementioned studies, most discussions about the relationship between tactile sensations and taste are predominantly focused on temperature variations and external sensory perceptions related to the weight and texture of tableware. To our knowledge, there is relatively less research on the influence of vibration in the oral cavity on taste.

## 2.2 Human Food Interaction

HFI constitutes a facet of human-technology interaction focused on supporting food-related practices, operating as a subset within the broader domain of HCI. In accordance with the insights of Khot and Mueller [37], the HFI realm encompasses two fundamental dimensions: the instrumental facet, which pertains to the utilitarian employment of technology in food-related contexts, such as the role of food in health and well-being; and the experiential dimension, which delves into the multisensory and identity-related experiences tied to food, encompassing pleasure, identity, and societal aspects. The conceptual framework presented by Khot and Mueller elucidates four distinctive phases encompassing HFI: cultivation, culinary preparation, consumption, and disposal. Based on our focus on the new values of modern food (such as discovery, experimentation, novelty, individuality, and inclusivity), and our aim to investigate the influence of cymatics on the dining experience, this section will provide a brief overview of the consumption category, with a specific emphasis on the influence of visual, auditory, and tactile aspects in the context of HFI applications.

Khot and Mueller [37] categorize HFI literature on eating into segments such as healthy eating, mindful eating, commensality (social eating), solo dining, and augmented eating. Within our study, augmented eating emerges as the salient focus due to its intrinsic relevance. Fukaike et al. [18] developed a fork integrated with a mechanism to release the aroma of allicin, providing controlled exposure to the nose for a comprehensive eating experience and effectively eliminating the occurrence of bad breath. FoodSkin [35] is an innovative technique for creating electronic circuits on the surfaces of food items utilizing edible gold leaf. This method enables the modulation of food taste through electrical stimulation, precise temperature control, and the provision of auditory feedback during the consumption of confections. When it comes to enhancing taste perception through electric circuits, Nakamura and Miyashita [60] have conducted extensive research in this area. Their studies have explored various utensils, such as straws, forks, and chopsticks, that can facilitate taste perception. Ueno et al. [93] also explored research related to improving the beverage experience using electrical stimulation. In the research on HFI regarding visual impact, there exists a correlation between dynamic visuals and gustatory perception. James et al. [30] unveiled that participants' flavour perceptions are enhanced by various video content, demonstrating substantial distinctions in their perceived taste sensations. These findings hold the potential to advance the concept of digital commensality and encourage healthier dietary practices. Miyashita [54] utilizes flavoured liquids to generate a printed image on the surface of different foods, transforming their appearance and taste into that of other food items. MetaCookie+ [62] and Arnold et al. [3] employ virtual reality to explore our dining experience. FoodFab [45] also utilizes food 3D printing to create perceptual illusions for controlling the level of perceived satiety. In addition to enhancing the dining experience through visual influence, there are also studies that employ a visual canvas to record gustatory perceptions. FlavourFrame [2] is a canvas-based app designed to assist tasters in capturing and visualizing their perceptions during mindful tasting experiences. The Vocktail [71], Taste+ [70], and FunRasa [69] system employ sensory modalities and electrical stimulation to generate virtual flavours and enhance the inherent flavours of a beverage. Upon consuming a beverage from the system, the integrated visual aspect (RGB light projected onto the beverage), taste component (electrical stimulation at the tongue's tip), and olfactory stimulus (emitted by micro air-pumps) synergize to engender a virtual flavour sensation, thereby modifying the beverage's taste profile. When it comes to research related to sound, numerous studies in the field of HFI suggest that technology and devices can be utilized to enhance the dining experience [10, 96]. For instance, in Wang et al. [101]'s study, "The Singing Carrot," designed an interactive device that combines sound with food, allowing users to experience multisensory stimulation while eating. Additionally, Chewing Jockey [40] augmented food texture by employing sound to enhance the quality of life for denture users. EducaTableware [33] mapped sounds during the eating and drinking process to promote healthier eating habits. Another example is Auditory Seasoning [38], a mobile app that offers various curated audio modes to alter chewing sounds. In the context of tactile impact, Hashimoto et al. [23] conceptualized a straw-like user interface to simulate a drinking experience virtually through vibrations. LOLLio [59], on the other hand, utilizes a gustatory interface by employing a lollipop as a tactile input device to gamify flavour shifts. In addition to straw-like user interface from Hashimoto et al. and LOLLio, which integrates tactile feedback with eating, Iwata et al. [27]'s Food Simulator also incorporates tactile technology to present food texture and chemical taste. Furthermore, Mesz et al. [49] also developed a wine glass that enhances the drinking experience through the combined effects of vibration and sound.

The aforementioned works are efficacious in addressing distinct issues and alleviating associated challenges. However, upon closer examination of augmented eating, it becomes apparent that many food technology devices necessitate supplementary equipment. Moreover, current research on improving the experience of beverages or liquid foods in the context of HFI often focuses on using electrical stimulation to enhance taste perception. While there are also studies exploring the influence of colour changes, added aromas, and even the incorporation of sound to enhance taste perception, there is relatively limited research on directly altering the shape of the liquid itself through physical principles. Additionally, even when utilizing vibration to enhance the texture of beverages, there is a scarcity of research that provides a clear understanding of the positive and negative impacts of HFI research. Furthermore, certain endeavours employ electrodes to enhance taste perception, a practice that has been highlighted as posing potential risks for children[7]. These limitations underscore the practical challenges in translating food tools devised within the HCI community into real-world applications. Therefore, we propose three hypotheses to investigate the impact of cymatics on taste perception and the dining experience. Furthermore, we aim to apply the research findings to beverage containers.

## 3 USER STUDY 1

In user study 1, our objective was to assess whether the hypothesized visual and tactile influences generated by cymatics could impact taste perception.

### 3.1 Taste Stimuli

In this phase, we prepared approximately 400ml samples of sweet, sour, bitter, salty, and umami solutions. The selection of taste stimuli, encompassing sweetness, saltiness, and umami, was guided by Wang et al.'s [100] cross-modal experiment on sound and taste perception, as well as Obrist et al.'s [65] research on taste experiences. For the sourness and bitterness tests, our choice of ingredients was influenced by Chamma et al. [11] and Ervina et al. [16], who utilized lemon and coffee (as well as Christina et al.'s [12] experiment on bitterness) in their respective studies. Ultimately, each of the five solutions was prepared with 400ml of drinking water and the following ingredients: 30ml of high-fructose corn syrup (Moriyama Sugar Syrup[8]) for the sweet solution, 6ml of 100% concentrated reduced lemon juice (Pokka Sapporo Lemon[9]) for the sour solution, 7g of highly roasted coffee (UCC Craftsman's Dark Roast Coffee Drip[10], based on the bitterness degrees of roasting described as "light," "medium," or "dark" in Buffo and Cardelli-Freire [9]) for the bitter solution, 5g of salt (NaCl, Salt Business Center[11]) for the salty solution, and 10g of monosodium glutamate (MSG, Ajinomoto Umami[12]) for the umami solution. The temperatures of these five solutions were maintained at room temperature. This precaution was taken to avoid potential influences on taste perception arising from temperature variations [88].

### 3.2 Cymatics Stimuli

Regarding the cymatics stimuli in user study 1, the phenomenon of cymatics is generated through auditory means [48]. Therefore, the selection of cymatics stimuli was inspired by a sound-matching task conducted at the Crossmodal Research Laboratory at the University of Oxford [100]. We mapped 10 frequencies on the tone generator, ranging from 27.5Hz, equivalent to the piano bass tone (Steinway

---

[7]https://gourmet.watch.impress.co.jp/docs/news/1438322.html
[8]https://www.fujimilk.co.jp/products/gum3.html
[9]https://www.pokkasapporo-fb.jp/english/about.html
[10]https://ucc-apac.com/en-ap/brands/ucc-craftsmans-coffee/
[11]https://www.shiojigyo.com/english/products/
[12]https://www.ajinomoto.com/brands/aji-no-moto

Table 1: Cymatics Frequencies

| Note Name | Frequency f(n) (Hz) | Volume (MacBook Pro) Nobsound Amplifier 100%, Szynalski Generator 75% |
| --- | --- | --- |
| A0 | 27.5Hz | Level 3 |
| D#1 | 38.8Hz | Level 2 |
| A1 | 55.0Hz | Level 2 |
| D#2 | 77.7Hz | Level 2 |
| A2 | 110.0Hz | Level 3 |
| D#3 | 155.5Hz | Level 4 |
| A3 | 220.0Hz | Level 5 |
| D4 | 293.6Hz | Level 6 |
| A4 | 440.0Hz | Level 7 |
| E5 | 659.2Hz | Level 10 |

C, A0) [19], to 659.2Hz, equivalent to piano E5 [51] (specifically: 27.5Hz, 38.8Hz, 55.0Hz, 77.7Hz, 110.0Hz, 155.5Hz, 220.0Hz, 293.6Hz, 440.0Hz, 659.2Hz). All frequencies were generated using the Szynalski tone generator[13] on a MacBook Pro running macOS Mojave. The volume was amplified using a Nobsound digital amplifier set at 100% volume level. The choice of volume level was determined by the visibility of the cymatics pattern, thus the range of volume levels controlled by the MacBook Pro (with the Szynalski tone generator set to 75%) spanned from level 2 to level 10, corresponding to different frequencies as detailed in Table 1 and Figure 2. As a speaker generating dynamic cymatic shape, we employed the Z-Modena MK2 as a drinking container. Given that colour can potentially influence taste perception [83], in user study 1, both the baseline initial mode and subsequent randomized taste tests consistently utilized the Z-Modena MK2 as the unchanging background colour for solutions. This approach aimed to minimize variables arising from colour variations throughout the study.

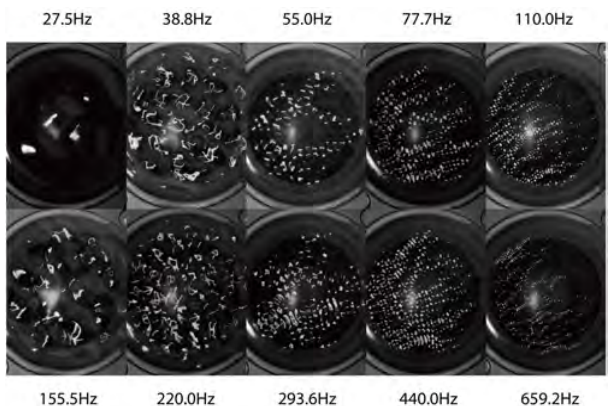

Figure 2: Cymatics pattern for tasting

[13]https://www.szynalski.com/tone-generator/

### 3.3 Participants

A total of 42 participants were enrolled in this study. For the user study 1, 20 participants (11 females, 9 males) aged between 18 and 34 years (M = 25.85, SD = 4.85) participated. Prior to their involvement, participants provided written informed consent and affirmed the absence of allergies to the ingredients and beverages used in the study (including corn syrup, lemon juice, coffee, salt, and MSG), as well as normal sensory capabilities in terms of visual, auditory, tactile, taste, and olfactory senses. Participants with cold symptoms were excluded from participation due to COVID-19 concerns. The study protocol was granted exempt approval by the university ethics review board. As a token of appreciation, each participant received a small dessert (such as cake, cookie, or other confectionery) upon completion of the study.

### 3.4 Procedure

This lasted approximately 90-120 minutes for each participant. As previously mentioned, cymatics involves three variables: visual, haptic, and auditory. Since our study specifically examines the effects of visual and tactile aspects of cymatics, participants were required to wear Audio-Technica ATH-M20x Professional Monitor Headphones and listen to Milli Hughes's Deep Phase Noise 1 at an iPhone 12 volume level of 8, which served as white noise in the background. This measure was taken to eliminate the influence of sound from the speaker.

Participants completed drinking tasks under various cymatics frequency testing conditions, each accompanied by their reported taste perceptions. Before conducting tests with random output frequencies, the initial mode (where all settings are standardized, including the use of the same headphones, speaker, straw, solution, and questions, with the only variation being the absence of cymatics effects) serves as the initial step in all experiments. All participants will provide ratings for the initial mode, serving as a baseline for comparison with subsequently randomly outputted frequencies. User study 1 was conducted at a laboratory environment and comprised two parts. The first part of the user study focused on the visual impact on taste perception, while the second part examined the tactile impact on taste perception.

In the first phase, participants were instructed to start by playing white noise and then take a sip of drinking water to refresh their palate before tasting any prepared solutions. For the sweetness test, participants used the speaker as their drinking vessel and a straw to consume a sweet solution. The procedure involved the following steps: 1) beginning with the initial mode, 2) evaluating their drinking experience, 3) taking a sip of drinking water to refresh their taste perception, and 4) consuming the sweet solution with 10 different frequencies presented randomly. Before each frequency change, participants assessed their drinking experience and were instructed to take another sip of water.

The questionnaire developed for participants is designed to evaluate their drinking experience. Upon participants' arrival in the experimental space, paper questionnaires are distributed. This survey utilizes a 9-point Likert scale, where the flavour intensity of each solution at different frequencies is assessed on a scale ranging from 1 to 9 (1 = not at all, 9 = extremely). Questions are formulated in the format: "Please evaluate your tasting experience: Sweetness," and participants are instructed to provide a numerical response between 1 and 9. Moreover, after completing the questionnaire, participants are given the opportunity to provide open comments. Following the first sweetness test, participants take a 5-minute break before progressing to the subsequent test, which includes evaluations for the remaining four flavour attributes (sourness, bitterness, saltiness, and umami).

To investigate the visual impact of cymatics on taste perception, a critical aspect of our first hypothesis, participants were instructed during the experiment to focus their visual attention on the liquid for approximately 10 seconds. They observed any shape changes induced by the cymatics pattern, subsequently using a straw to drink from the solution without touching the surface of the speaker to minimize any potential tactile influence. The entire experimental process is depicted in Figure 3.

In the second phase, our attention shifted towards investigating the tactile impact of cymatics on taste perception, a pivotal aspect of our second hypothesis. Differing from the first phase, participants were instructed to close their eyes to mitigate visual influences. They were asked to touch the surface of the speaker with the straw, feeling the vibrations for approximately 10 seconds before proceeding to drink the solution. The remaining procedural aspects of this phase remained consistent with those of the first phase.

Figure 3: The process of user study 1

## 3.5 Results

Based on the following data, H1 has been supported in user study 1. Statistical software SPSS 29.0.1 (SPSS, Inc.) for Mac was employed for result analysis. We conducted a one-way ANOVA analysis with taste perception (sweetness, sourness, bitterness, saltiness, and umami) and cymatics frequencies as the factors affecting participants' ratings. Pearson correlation was utilized to assess the relationships between different frequencies, and independent-samples t-tests were performed when comparing two factors was necessary for additional verification.

### 3.5.1 The visual impact generated by cymatics on taste perception.
Figure 4 (a)-(e) illustrates the ratings of sweetness, sourness, saltiness, umami, and bitterness for ten different visual frequencies (27.5 Hz, 38.8 Hz, 55.0 Hz, 77.7 Hz, 110.0 Hz, 155.5 Hz, 220.0 Hz, 293.6 Hz, 440.0 Hz, 659.2 Hz). Through one-way ANOVA, we observed a significant overall effect of visual frequencies on bitterness ratings ($F(10, 209) = 2.064, p = 0.029, \eta^2 = 0.090$). In the case of bitterness, Pearson correlation coefficients (r) were calculated to evaluate the correlations among the different visual frequencies. The results revealed significant positive correlations between 27.5 Hz and 77.7 Hz ($r = 0.78, p < 0.001$), 27.5 Hz and 110.0 Hz ($r = 0.72, p < 0.001$), 77.7 Hz and 293.6 Hz ($r = 0.78, p < 0.001$), 77.7 Hz and 440.0 Hz ($r = 0.71, p < 0.001$), 110.0 Hz and 220.0 Hz ($r = 0.80, p < 0.001$), 110.0 Hz and 293.6 Hz ($r = 0.80, p < 0.001$), 155.5 Hz and 440.0 Hz ($r = 0.75, p < 0.001$), and 293.6 Hz and 440.0 Hz ($r = 0.67, p < 0.001$). In general, when considering the impact of visual frequencies, bitterness exhibited a significantly higher effect compared to the other four taste modalities. Additionally, visual frequencies of 27.5 Hz resulted in a significant perception of bitterness compared to other frequencies.

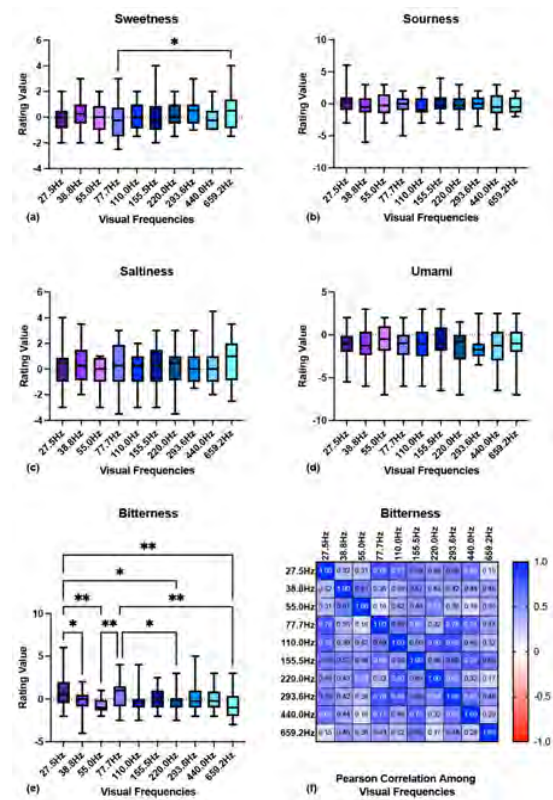

Figure 4: (a)-(e) Flavour ratings influenced by visual impact generated through cymatics. Asterisks denote significance levels calculated via one-way ANOVA: * $p < .033$, ** $p < .002$, *** $p < .001$ (f) Pearson correlation among visual frequencies in bitterness testing

### 3.5.2 The tactile impact generated by cymatics on taste perception.
Figure 5 (a)-(e) illustrates the ratings of sweetness, sourness, saltiness, umami, and bitterness for ten different tactile frequencies (27.5 Hz, 38.8 Hz, 55.0 Hz, 77.7 Hz, 110.0 Hz, 155.5 Hz, 220.0 Hz, 293.6 Hz, 440.0 Hz, 659.2 Hz). We did not observe a significant main effect of tactile frequencies via one-way ANOVA tests on sweetness ($F(10, 209) = 0.695, p = 0.728$), sourness ($F(10, 209) = 0.275, p = 0.986$), saltiness ($F(10, 209) = 0.436, p = 0.928$), umami ($F(10, 209) = 1.549, p = 0.124$), and bitterness ($F(10, 209) = 0.633, p =$

0.785). Therefore, no further follow-up Pearson correlation tests were conducted.

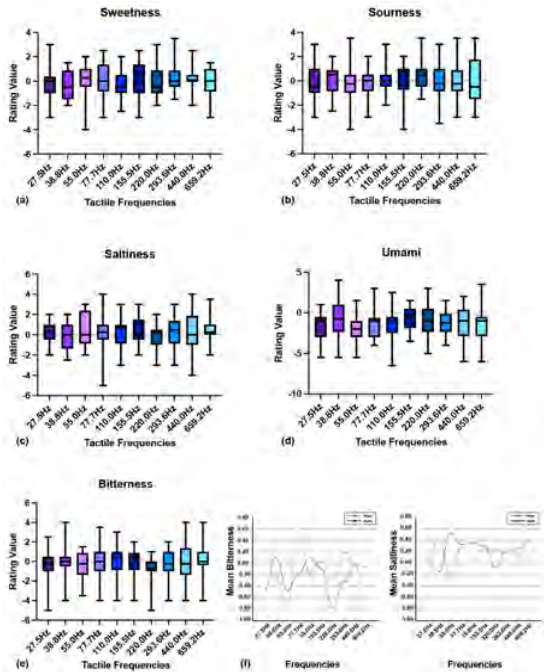

Figure 5: (a)-(e) Flavour ratings influenced by tactile impact generated through cymatics (f) The correlation between visual frequencies and tactile frequencies

*3.5.3 The correlation between visual and tactile influences on taste perception.* Figure 5 (f) depicts the correlation between visual frequencies and tactile frequencies. A significant effect of 27.5 Hz on bitterness ratings was observed, as indicated by a one-way ANOVA ($F(1, 38) = 4.193, p = 0.048, \eta^2 = 0.099$). Additionally, a t-test was conducted, revealing significance in saltiness rating ($t(38) = 0.471, p = 0.034, 95\%CI[-0.74218, 1.1918]$).

*3.5.4 Qualitative Analysis.* In user study 1, we conducted a qualitative analysis of 13 spontaneously provided open comments to elucidate the influence of visual and tactile impact on taste perception. The transcribed transcripts served as the foundation for an open thematic coding process.

**Visual and tactile impact (7 cases, ≈ 53.85%).** P1 expressed observations such as, "*When referring to the visual impact on umami testing, the impact is more noticeable on the tip of the tongue, but essentially the same at the base of the tongue.*" This aligns with the overall conclusion of the theme concerning the comparison between visual and tactile impact. P2 also noted, "*It's challenging to discern taste during the visual test for saltiness,*" validating the non-significant analysis result for saltiness testing. P1, P2, P3 and P4 unanimously expressed, "*Vibration has a minimal effect on taste, and it even feels like an additional source of interference, disrupting the tasting experience,*" and "*The stronger the vibration, the weaker the taste.*" However, P5's feedback differed: "*During the bitterness vibration test, I felt the taste getting stronger each time.*" Based on the feedback from participants P1, P2, P3, and P4, we observed differing perspectives compared to previous research that highlighted the positive impact of vibration on dining experiences [23, 49, 59].

**Comparison between visual and tactile impact (3 cases, ≈ 23.08%).** This theme received feedback consistent with the visual and tactile impact theme, P6 and P7 noted, "*Differences caused by visual impact are more pronounced than those caused by vibration.*" This observation further supports the influence of visual impact on taste, as previously discussed [75, 92]. Additionally, P8 remarked, "*During umami testing, the drink is more enjoyable in the visual experiment, but less enjoyable in the vibration experiment.*"

**Eating habits and dining experience (3 cases, ≈ 23.08%).** P9 expressed, "*Because I frequently consume coffee, I may not perceive bitterness well and am less sensitive to it. Therefore, individuals with different cultural and dietary habits may yield different results.*" P10 shared, "*I rarely consume foods with bitter or umami flavours in daily life, so I find it challenging to provide ratings for these two tastes.*" These observations align with Tokat et al.'s [91] view of demographic structure influences. P8 stated, "*Regardless of whether it's the visual or vibration experiment, the higher the frequency, the sweeter it tastes.*" This observation corresponds with previous research on the relationship between sound and taste [100]. Further discussion of these findings will be presented in the subsequent Discussion section.

## 4 USER STUDY 2

In user study 2, our objective was to investigate the third hypothesis: whether the overall influence of cymatics could enhance the drinking experience, including factors such as texture, appetite, and enjoyment. To determine whether the effects of cymatics on the drinking experience were solely due to its inherent characteristics (dynamic deformation and vibration), or if specially designed cymatics patterns were required for an enhanced overall experience, we categorized the cymatics stimuli setup into two distinct categories, as elaborated in the following section.

### 4.1 Taste Stimuli

In this phase, we opted for commonly consumed beverages to assess the practicality of the concept. A 580ml Bomy guava juice[14] containing 58g of sugar was chosen for the sweetness test to highlight the effect. In the study by Yonis et al. [107], guava juice not only emits a pleasant sweet refreshing aroma but also, because Bomy's guava juice contains 10% sugar, Lichtenstein and P. E. [44] indicated that at a sucrose concentration of 10%, dextrose has a relative sweetness value of 65. This served as the rationale for selecting Bomy guava juice. To evaluate sourness, we opted for a lemonade[15] made with water, sugar, and concentrated reduced lemon juice, as the enhanced perception of sourness has been observed with lemon extract [98]. Additionally, lemonade was used as an experimental case in the HFI study by Ranasinghe et al. [68]. For bitterness evaluation, Chinese holly leaf (Folium Ilicis Latifoliae)[16] was selected.

---
[14] http://en.bomy.com.tw/juice/
[15] https://foodservice.weichuan.tw/products/冷藏農搾檸檬飲_900ml
[16] https://www.coteahouse.com.tw/product2.html

This ingredient, recognized in traditional Chinese medicine [108], contains triterpene glycosides, sterols, and bitter compounds [110], leading to its inclusion. Due to its recognition as a medicinal herb with high bitterness content and its potential use in fighting against COVID-19 [43], its bitter nature also allows us to test the efficacy of cymatics when using ingredients that are generally disliked by the public. Japanese miso soup[17] was chosen for the salty test, inspired by the research on Taste-Adjusting Chopsticks[18] (Miyashita et al., 2023 [55]) conducted by Miyashita Laboratory and Kirin. In experiments where electrical currents affect saltiness, miso soup has been widely used and effectively validated in previous studies [34, 72]. Also, because miso soup is commonly used as one of the foods to judge salt intake in many studies [87, 94], these considerations led to our selection of miso soup. Finally, for umami assessment, seaweed soup[19] was selected, specifically utilizing brown seaweed kombu (Saccharina japonica), known for its high free glutamate content, commonly used in Japanese soup stock dashi [58]. Additionally, seaweed has been widely used as a food case study for investigating umami in many studies [17, 52, 58], making seaweed soup our preferred choice for the umami drinking experience. The temperatures of these five ingredients were maintained at room temperature. This means that hot soups, such as miso soup and seaweed soup, were allowed to cool to room temperature before the experiments began (the preparation of soups followed the official product instructions). This precaution was taken to avoid potential influences on taste perception arising from temperature variations [88].

### 4.2 Cymatics Stimuli

As mentioned earlier, in this study, we categorized cymatics stimuli into two groups to test whether general cymatics effects alone could enhance the drinking experience or if specially designed cymatics were required to influence the drinking experience. First, let's introduce the cymatics designed based on the results of user study 1, and then we'll discuss general stimuli capable of presenting cymatics effects.

Building upon the previous user study's results for the cymatics stimuli, in user study 2, we established the corresponding relationships between the visual and tactile effects generated by cymatics and the five taste perceptions. The cymatics stimuli for each taste were created by building a sonification system within Max/MSP software. In this system, each sound had its own designated frequencies, along with their designated amplitude levels. These data were obtained from user study 1, specifically the frequencies selected by previous participants. The Max/MSP software took these frequency and amplitude limitations and generated rhythms using a semi-randomized approach by manipulating the envelopes of each sound and the frequency of their playback. The sound artist played these sounds back on the device until a desired aesthetic outcome emerged from the randomized elements. These compositions, which we refer to as "Cymatics Seasoning," corresponded to the following taste categories: sweet cymatics (293.6Hz and 659.2Hz), sour cymatics (155.5Hz and 220.0Hz), bitter cymatics (27.5Hz, 38.8Hz, and 110.0Hz), salty cymatics (55.0Hz and 659.2Hz), and umami cymatics (155.5Hz). Further details regarding the specific cymatics compositions for each flavour are summarized in Table 2 and Figure 6.

Table 2: Cymatics Seasoning for each flavour

| Flavour | Frequencies |
| --- | --- |
| Sweet | 293.6Hz, 659.2Hz |
| Sour | 155.5Hz, 220.0Hz |
| Bitter | 27.5Hz, 38.8Hz, 110.0Hz |
| Salty | 55.0Hz, 659.2Hz |
| Umami | 155.5Hz |

In addition to the five flavour-Cymatics Seasoning we had prepared, we conducted an experiment to investigate whether the application of cymatics patterns would still impact the drinking experience when using general pop music. For this purpose, we selected JHM and Fractures' "Fever Dream"[20] due to its consistent low-pitched tones throughout the composition, which allowed for the sustained generation of cymatics patterns during the whole process of user study 2.

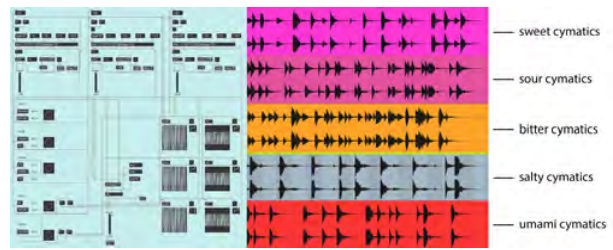

Figure 6: Five different Cymatics Seasoning rhythms corresponding to sweet, sour, bitter, salty, and umami

### 4.3 Participants

In user study 2, we recruited 22 participants (15 females, 7 males) aged between 24 and 63 years (M = 30.22, SD = 9.96). Prior to their involvement, participants provided written informed consent and affirmed the absence of allergies to the ingredients and beverages used in the study (including guava juice, Chinese holly leaf, miso soup, and seaweed soup), as well as normal sensory capabilities in terms of visual, auditory, tactile, taste, and olfactory senses. Participants with cold symptoms were excluded from participation due to COVID-19 concerns. The study protocol was granted exempt approval by the university ethics review board. As a token of appreciation, each participant received a small dessert (such as cake, cookie, or other confectionery) upon completion of the study.

### 4.4 Procedure

During user study 2, in contrast to individual experiments, a group session was conducted with approximately 90 minutes duration,

---

[17]https://www.marukome.co.jp/product/detail/instant_100/
[18]https://www.kirinholdings.com/en/newsroom/release/2022/0411_01.html
[19]https://www.orionjako.com/複製-トッポギチャプチェ

[20]https://youtu.be/lwjj_RnWdcc?si=OgS_gzHu2tshO8oc

involving 22 participants simultaneously. Unlike the previous individual user study where participants wore headphones and were situated in a laboratory setting, user study 2 was held in a cafeteria without the use of additional wearable devices. This setup aimed to observe the real-world impact of cymatics on participants' experiences. Participants were guided through a series of drinking tasks involving both cymatics pop music and various Cymatics Seasoning, which were developed based on the user study 1 results. The objective was to investigate our third hypothesis, which focused on the influence of cymatics on appetite and overall enjoyment of the drinking experience. Before conducting the drinking experience impact tests with Cymatics Seasoning and cymatics generated by pop music, participants initiated with an initial mode (using the same questionnaire and taste stimuli but without any cymatics effects), using a standard 100ml paper cup as the first step in all experiments. All participants will provide ratings for the initial mode, serving as a baseline for comparison with subsequently outputted Cymatics Seasoning and cymatics pop music.

The procedure for the sweetness test consisted of the following steps: 1) taking a sip of drinking water to refresh taste perception, 2) tasting the guava juice with the initial mode, 3) rating their drinking experience, 4) taking another sip of drinking water, 5) tasting the guava juice with a speaker playing the sweet cymatics, 6) rating their drinking experience once more, 7) drinking another sip of water, 8) tasting the guava juice with a speaker playing JHM and Fractures' "Fever Dream", and 9) providing a final rating of their drinking experience. After completing the sweetness test, participants were given a 5-minute break before proceeding to the remaining four flavour tests (sourness, bitterness, saltiness, and umami), each separated by a 5-minute break.

The questionnaire designed for participants aims to evaluate their drinking experience. Upon participants' entry into the experimental space (cafeteria), paper questionnaires are distributed. The questionnaire used in user study 2 drew inspiration from Kleinberger et al. [38], who studied the influence of chewing sounds on food augmentation. This survey assessed participants' drinking experience through 10 questions, including the flavour intensity of sweetness (adjusted for other flavours such as sourness, bitterness, saltiness, or umami in the case of other flavour tests), overall flavour intensity, overall deliciousness, mouth-filling sensation, level of fullness after drinking, desire to take another sip, attention to the beverage in general, attention to the flavour of the beverage, attention to the texture of the beverage, and the overall level of enjoyment. The questionnaire uses a 9-point Likert scale to assess the experience intensity of each solution under the influence of two different cymatics effects (Cymatics Seasoning and cymatics generated by pop music). The scoring range is from 1 to 9 (1 = not at all, 9 = extremely). The question format is "Please evaluate your tasting experience: Overall level of enjoyment," and participants are asked to provide a numerical response between 1 and 9. Additionally, after completing the questionnaire, participants can choose whether to provide open comments. Open comments are divided into three topics: "Do you wish us to develop customized frequency seasonings in our future projects? Why?", "How did the cymatics change your drinking behavior/experience?", and "Any feedback you may have."

## 4.5 Results

Based on the following data, H3 has been supported in user study 2. Statistical software SPSS 29.0.1 (SPSS, Inc.) for Mac was employed for result analysis. We conducted a one-way ANOVA analysis with taste perception (sweetness, sourness, bitterness, saltiness, and umami), Cymatics Seasoning, drinking experience, and cymatics generated by pop music as the factors affecting participants' ratings. Independent-samples t tests were performed when comparing two factors was necessary for additional verification.

### 4.5.1 The Impact of Cymatics Seasoning on the Drinking Experience.
Figure 7-8 illustrate the ratings given to guava juice, lemonade, Chinese holly leaf (Folium Ilicis Latifoliae), miso soup, and seaweed soup across ten different drinking experiences with Cymatics Seasoning and cymatics generated by pop music. Through one-way ANOVA, we identified a significant overall effect of Cymatics Seasoning on the ratings of these five beverages/soups: guava juice ($F(9, 210) = 5.914, p < 0.001, \eta^2 = 0.202$), lemonade ($F(9, 210) = 2.992, p = 0.002, \eta^2 = 0.113$), Chinese holly leaf ($F(9, 210) = 27.384, p < 0.001, \eta^2 = 0.539$), miso soup ($F(9, 210) = 4.103, p < 0.001, \eta^2 = 0.149$), and seaweed soup ($F(9, 210) = 5.771, p < 0.001, \eta^2 = 0.198$).

In comparison to the baseline, subsequent one-way ANOVA revealed the significance of sweet flavour intensity ($F(1, 42) = 7.865, p = 0.008$) and general attention ($F(1, 42) = 4.478, p = 0.040$) for guava juice, as well as umami flavour intensity ($F(1, 42) = 4.954, p = 0.031$), overall flavour intensity ($F(1, 42) = 5.818, p = 0.036$), and flavour attention ($F(1, 42) = 4.122, p = 0.049$) for seaweed soup. Furthermore, a t-test was conducted, revealing significance in sweet flavour intensity rating ($t(42) = 2.804, p = 0.044$, 95% CI [0.32416, 0.25492]) for guava juice, overall deliciousness rating ($t(42) = 0.775, p = 0.048$, 95% CI [-0.58333, 1.31061]), and enjoyment rating ($t(42) = 0.793, p = 0.043$, 95% CI [-0.56194, 1.28921]) for lemonade.

### 4.5.2 The Impact of Cymatics Generated by Pop Music on the Drinking Experience.
Figure 7-8 present the ratings given to guava juice, lemonade, Chinese holly leaf (Folium Ilicis Latifoliae), miso soup, and seaweed soup across ten different drinking experiences with cymatics generated by pop music and Cymatics Seasoning. Through one-way ANOVA, a significant overall effect of cymatics pop music on the ratings of these five beverages/soups was observed: guava juice ($F(9, 210) = 2.857, p = 0.003, \eta^2 = 0.109$), lemonade ($F(9, 210) = 2.673, p = 0.006, \eta^2 = 0.102$), Chinese holly leaf ($F(9, 210) = 21.506, p < 0.001, \eta^2 = 0.479$), miso soup ($F(9, 210) = 3.549, p < 0.001, \eta^2 = 0.132$), and seaweed soup ($F(9, 210) = 5.663, p < 0.001, \eta^2 = 0.195$).

In comparison to the baseline, subsequent one-way ANOVA revealed the significance of general attention ($F(1, 42) = 4.653, p = 0.037$) for guava juice. Additionally, a t-test was conducted, revealing significance in the fullness level rating ($t(42) = 1.154, p = 0.050$, 95% CI [-0.51068, 1.87432]) for guava juice.

### 4.5.3 The Correlation Between Cymatics Seasoning and Cymatics Pop Music on the Drinking Experience.
Through one-way ANOVA tests, we did not observe a significant main effect of the correlation between Cymatics Seasoning and cymatics pop music on guava juice ($F(1, 438) = 0.044, p = 0.834$), lemonade ($F(1, 438) = $

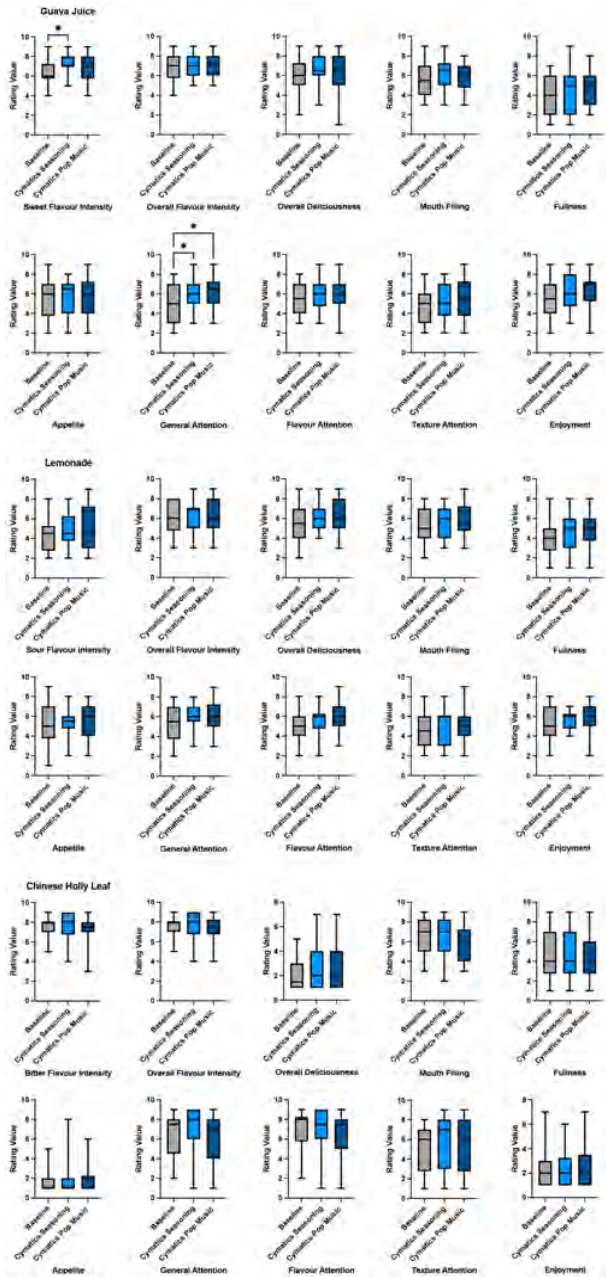

Figure 7: Drinking experience ratings influenced by Cymatics Seasoning and cymatics pop music on guava juice, lemonade, and Chinese holly leaf

0.629, $p = 0.428$), Chinese holly leaf ($F(1, 438) = 0.928, p = 0.336$), miso soup ($F(1, 438) = 0.909, p = 0.370$), and seaweed soup ($F(1, 438) = 2.952, p = 0.086$). Therefore, no further follow-up independent-samples t tests were conducted.

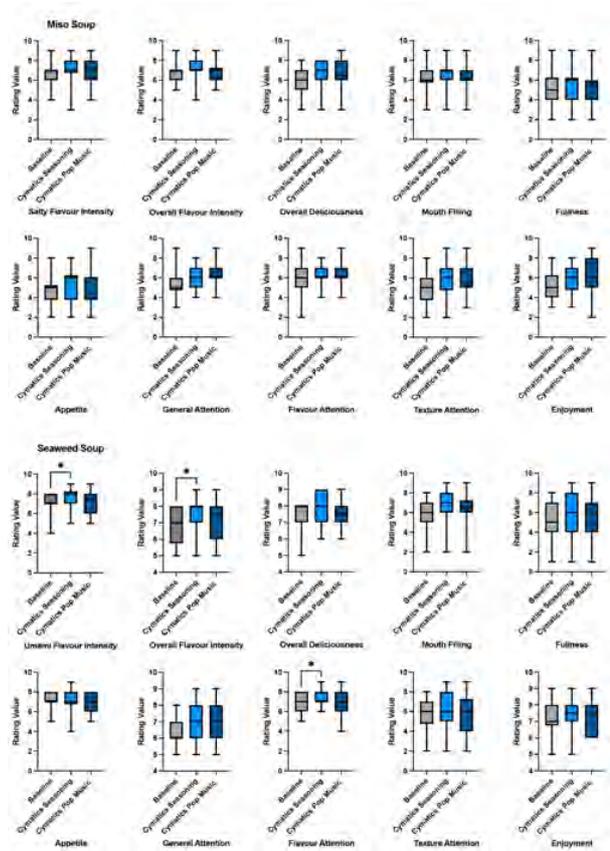

Figure 8: Drinking experience ratings influenced by Cymatics Seasoning and cymatics pop music on miso soup and seaweed soup

*4.5.4 Qualitative Analysis.* Transcriptions were made for all 71 open comments, and a qualitative analysis was performed using the transcribed data. In user study 2, we included two specific questions and a final open-ended comment. Here, we structured our analysis around three different questions, employing an open thematic coding process. These three questions are: "Do you wish us to develop customized frequency seasonings in our future projects? Why?", "How did the cymatics change your drinking behavior/experience?", and "Any feedback you may have."

**Development suggestions (15 cases, ≈ 21.13%).** Excluding comments with limited detail such as "*looking forward to it,*" participants actively provided specific suggestions for future development. Examples include P1 "*It could be included colour changes for dining experiences,*" P3, P9, and P21 "*Hope to have the function to freely adjust patterns,*" P12 "*It could be integrated with the catering industry, where decoration and background music may also affect the experience,*" and P14 "*Future research could include detection of brainwaves,*" P19 "*In terms of artistic creation, it could be collaborated with sound art or participatory art.*" In addition to suggestions for diversifying sensory experiences, such as introducing the influence of colour [83], designing dining space [80], coordinating background music [24, 50, 73, 84, 99], exploring brainwave research, and integrating with sound art and participatory art, several participants also noted potential health applications. Examples include P17 and P19 "*Relying on cymatics for seasoning might help patients with taste disorders. It could be used in medicine or help the elderly avoid consuming heavily seasoned foods harmful to their health.*" However, there are still 3 cases indicating negative comments for future development. For instance, P7 and P11 "*It's okay, the effect doesn't seem very noticeable.*" Notably, P16 provided practical considerations, stating, "*I think this is a marketing direction, but without confirming how long the seasoning effect of cymatics can last, I think the development cost would be too high.*" This also suggested a new direction for future research: How long does the influence of Cymatics Seasoning last?

**Experiential and behavioral changes (38 cases, ≈ 53.52%).** P2, P6, P8, P10, P15, P18, and P20 consist of repetitive statements such as "*making eating more enjoyable,*" and "*highly appealing.*" However, P22 specifically noted, "*This can enhance the sophistication, uniqueness, and personalization of daily life.*" According to attention changes, most comments were highly repetitive, such as P1, P3, P4, P6, P9, and P19 "*enhances focus while drinking,*" "*will observe the liquid ripples up close and be surprised by treating the speaker as a drinking vessel,*" and similar expressions. When it comes to texture changes, P3, P4, and P9 "*will especially feel the texture of the drink,*" with only P5 explicitly stating, "*never thought cymatics would affect texture; bitterness becomes so strong that I don't want to take another sip.*" This comment aligns with the results of user study 1, indicating a significant impact of cymatics on bitterness. There are also some comments focused on flavour changes, with P2, P13, P17, and P18 "*obviously changes the taste, curious about the underlying principles,*" P19 "*The taste changes of umami are particularly noticeable.*" However, there were also more conservative comments, P14 and P15 stated "*slightly enhances or diminishes certain flavours.*" There are 10 cases focused on taste sensitivity/multi-sensory experience. P1, P6, P8, P9, P10, P21, and P22 positively expressed that "*increases the sensory experience and taste sensitivity,*" and "*drinking experience changes with the rhythm of music, creating new mental images or imaginings.*" However, P12 and P16 mentioned, "*compared to the baseline, the drinking experience and taste are different. However, drinking behavior is only turning the drink container into a speaker, and there seems to be no significant difference in post-behavior.*" According to the dining atmosphere, P7, P20 and P22 stated, "*mainly changes the atmosphere, making my experience different.*" While only 3 participants mentioned "dining atmosphere," it was relatively unexpected, yet it also corroborates previous research indicating that the atmosphere can influence the dining experience [28, 29, 41, 102].

**Experimental suggestions/user experience (18 cases, ≈ 25.35%).** P1 and P3 desire for "*continuous audio frequency output.*" P1 also provided specific recommendations for Cymatics Seasoning: "*During the Cymatics Seasoning testing phase, variations in audio tempo, rhythm, meter, vibrato, staccato, and tonality may yield different effects on the overall taste experience.*" Regarding these frequency output [78] suggestions for Cymatics Seasoning, we will incorporate them into future work. Concerning ingredient suggestions, P2 and P19 expressed that "*the bitterness itself is too strong, making it difficult to discern differences before and after the experiment.*" This indicates that in future research, milder-flavoured beverages should be considered to facilitate participants in distinguishing taste differences before and after the experiment. Additionally, P13 suggested, "*I recommend using the following for taste testing: sour: H+ concentration, sweet: (Acid-Base theory), salty: sodium concentration, umami: MSG+IMP*" While these comments provide valuable insights, they may be more applicable in experiments with controlled variables, unlike user study 2, which aimed to evaluate experiences in real dining situations. Some participants have opinions about the consumption method; P14, P15, P18 and P22 expressed, "*I think changing the taste might also be related to how you drink it, such as shortening the straw, or drinking it directly.*" This perspective aligns with Harrar and Spence's [22] view regarding the potential impact of different consumption methods on the dining experience. Additionally, P20 even suggested "*blindfolded tasting,*" although the practicality of removing visual influences in taste evaluation has already been explored in user study 1. P4 and P8 also express concern about the volume splatter experience, stated, "*Be careful about beverage splattering*." Future work will consider this aspect, particularly concerning the unintentional changes in audio volume by participants. The last aspect pertains to taste-related considerations. Excluding comments with limited information, such as those from P16 and P21 expressing that the experience is "*unique,*" P19 mentioned, "*During each taste, the third-stage cymatics pop music introduces additional flavours beyond the taste being explored, making it feel quite refreshing!*" This comment may lead to unexpected discoveries and holds potential research value. Additionally, despite instructions for participants to drink a small amount of water to cleanse their palate between each test and a 5-minute interval between each beverage, P17 expressed, "*I'm worried about taste fatigue.*" Taste fatigue remains a significant issue that requires optimization in future work. Lastly, P1 mentioned, "*It's challenging to distinguish the sensation of fullness.*" Based on this comment, future work may consider a standardized requirement for participants not to drink or eat one hour before attending the study [65].

## 5 DISCUSSION
### 5.1 Interpretation of Findings

*5.1.1 User Study 1.* As observed in user study 1, the 27.5Hz frequency's shape exerted a significantly positive influence on the perception of bitterness. Our initial assumption was grounded in the common association of carbonated, bitter, salty, and sour-tasting foods and drinks with angular attributes on the sensory scale. In contrast, sweet and creamy sensations are consistently linked with rounder shapes [80]. Therefore, we hypothesized that people would associate higher frequencies with bitterness due to the fragmented appearance of higher-frequency shapes, compared to the rounder shapes typically related to lower frequencies on the sensory scale. However, the results indicated that the 27.5Hz frequency's shape can enhance the perception of bitterness, aligning with a prior study suggesting a correlation between low pitch and bitterness [100]. Initially, our efforts were focused on minimizing the impact of auditory influence on taste perception [38, 49, 100, 101] during user study 1. Additionally, 27.5Hz frequency is categorized as a low-frequency sound, aligning with the findings of Møller et al. [56]. As the frequency decreases, human hearing becomes gradually less sensitive [56]. Despite these circumstances, we still observed a significant effect of 27.5Hz on bitterness, underscoring the substantial influence of sound on taste perception (because the phenomenon of cymatics is generated by sound). Alternatively, we can consider the relationship between the cymatics effect and the exceptional study conducted among the Himba tribe in Kaokoland, rural Namibia. In their case, milk chocolate is better matched with more angular shapes, while dark chocolate is associated with rounder shapes [8]. Another possibility is suggested by Deroy et al.'s research [15], where irregular edges were more frequently linked to a more bitter and sparkling flavour. Similarly, Turoman et al.'s study [92] confirmed that a lower number of reflection/rotation axes was linked to higher perceived sourness, bitterness, and threat of round shapes. In terms of visual perception, compared to other higher frequencies, the rhythmic sensation of 27.5Hz is relatively high. Therefore, the presentation of shapes at this frequency, while appearing more rounded than high-frequency shapes, is simultaneously more irregular and has lower reflectance. These findings may serve as evidence supporting their correlation.

Furthermore, there are noteworthy findings when comparing the visual and tactile impacts generated by cymatics. During the open feedback session, 83% of participants voluntarily expressed that the visual impact was more pronounced than the tactile impact. Some participants even reported that the vibrations from the speaker diminished their flavour sensation. Only one participant reported that "*During the bitterness vibration test, I felt the taste getting stronger each time.*"

Previous research has indicated positive effects of vibrations on dining experiences [49, 90], particularly in simulating virtual drinking sensations using pressure, vibration, and sound [23]. Moreover, previous research has indicated that the physical properties of vibration can result in a slight increase in acidity in red wine [13] and impact the texture of cheese sauce [74]. However, there is relatively limited exploration of the specific effects of vibration in crossmodal studies involving five taste solutions (sweet, sour, bitter, salty, umami). In our study, participants' voluntary feedback suggests that vibrations generated at specific frequencies may diminish taste perception. This phenomenon raises questions about which types of vibration can enhance the drinking experience and which cannot. In this study, we have demonstrated that the vibrations generated by cymatics can even diminish the drinking experience. Nonetheless, these results present opportunities for various potential applications. In the design of Cymatics Seasoning, we can explore different patterns and shapes in the visual aspect to enhance taste perception and the overall drinking experience. It is also important to consider and mitigate the potential influence of vibrations, recognizing that the effects of vibrations from specific frequencies may interfere with taste perception.

*5.1.2 User Study 2.* Based on the results of user study 2, we have come to understand the necessity of designing Cymatics Seasoning. We compared the differences between Cymatics Seasoning generated based on previous research data and general cymatics pop music and found that Cymatics Seasoning exhibited higher significance than cymatics pop music. During the testing phase of Cymatics Seasoning, in contrast to the baseline, sweetness, sourness, and umami positively influenced taste perception and the drinking experience, whereas, during the testing phase of cymatics pop music, only sweetness showed significant positive effects. This indicates that Cymatics Seasoning can be further explored and developed in the future. It can effectively enhance flavour intensity, attention, deliciousness, and enjoyment, aligning with our original intentions. It not only positively impacts taste perception but also has a strong connection with the new values of modern food.

It is worth noting that, unlike the results from user study 1, in the bitterness test of user study 2, bitterness showed significance only in the overall ratings, but there were no significant differences compared to the baseline, whether under the influence of Cymatics Seasoning or general cymatics pop music. This could be attributed to the close relationship between the selected ingredients. In the bitterness test of user study 2, we used Chinese holly leaf, which is considered an extremely bitter traditional Chinese medicinal herb. This also demonstrates that the influence of Cymatics is not strong enough to enhance the taste of ingredients that are already intensely flavoured or to diminish the undesirable tastes that are generally disliked by the public. This, in turn, might not make individuals more willing to consume liquid food they dislike. Further elaboration on this aspect is provided in the Limitations section.

Furthermore, in user study 2, when participants were asked about the potential future developments of the Cymatics Cup, many expressed positive sentiments and a strong desire to have the ability to customize Cymatics Seasonings. This might be associated with the tendency for objects that are closely related to oneself or, in some sense, belong to oneself, to be prioritized [85]. Spence's work [80] also mentions the self-prioritization effect, which helps explain why beverages often taste better when consumed from a favored mug. From these participant comments, we identify an opportunity to develop a comprehensive Cymatics Seasoning system. Such a system would empower individuals to create personalized beverages that encompass health, enjoyment, and appetite, all within a single cup.

In another set of open-ended comments, we gathered several valuable suggestions, including potential applications in the medical field for individuals with impaired taste and for promoting healthier eating among the elderly. Behrns et al. [5] identified a high demand for a liquid diet among hospitalized patients. Spence [81] highlighted the unfavorable experience of consuming hospital food. Therefore, given the potential of Cymatics Seasoning to enhance the dining experience, there is a significant opportunity for its application and development in the medical field.

Further insights from participants suggest the potential for more in-depth exploration of Cymatics Seasoning, such as introducing additional variables like tempo, rhythm, meter, vibrato, staccato, and tonality. Moreover, drawing on previous research on the relationship between music and taste [78], it can be affirmed that these variables could be crucial factors influencing the dining experience. Some participants also suggested experimenting with different consumption methods, such as shortening the straw length or directly drinking from the cup. However, due to the observation in user study 1 that vibration diminishes taste perception, the recommendations for changing consumption methods are not currently advised. This is because both shortening the straw and direct drinking may indirectly increase the perception of vibration. In our future work, efforts should be directed toward minimizing the impact of vibration on oral sensations. Nonetheless, the exploration of varied consumption methods can serve as a potential avenue for future research, as different drinking techniques may yield diverse dining experiences. In summary, these valuable insights gleaned from open-ended comments will be taken into consideration in our upcoming studies on Cymatics Seasoning and the Cymatics Cup.

In conclusion, insights from user study reveal the following:

(1) The visual enhancement of bitterness perception at 27.5Hz supports H1. Despite deviating from our initial expectations, subsequent findings align with literature supporting an association between irregular and asymmetrical shapes with bitterness, validating our discoveries.
(2) Vibration's potential to diminish taste perception supports H2. In contrast to prior research in the HFI field, which predominantly indicated positive effects of vibration on dining experiences, feedback from our participants mostly suggests limited positive impacts of vibration on taste perception.
(3) The creation of Cymatics Seasoning based on user study 1 results supports H3. In user study 2, it significantly enhanced sweet flavour intensity and general attention for sweet beverages, the deliciousness and enjoyment for sour beverages, and the umami flavour intensity, overall flavour intensity, and flavour attention for umami soup.
(4) While Cymatics Seasoning offers benefits, it cannot alter the taste of inherently strong-flavoured ingredients.
(5) The user experience of splashing during consumption and difficulty in identifying the speaker as the beverage vessel were noted for potential enhancements.

## 5.2 Prototype Design: The Vessel for Cymatics Tasting

To assess the influence of various cymatics on taste perception and drinking behaviours, both user study 1 and user study 2 utilized the Z-Modena MK2 as a drinking container. The Z-Modena MK2 features a 3.1-inch (8 cm) full-range speaker unit, as illustrated in Figure 9. Subsequently, following the positive results, a dedicated Cymatics Cup prototype was designed.

Given the outcomes of the user study, where participants initially expressed surprise at the concept of utilizing a speaker as a beverage container, it is imperative in the design of the Cymatics Cup to consider the user experience. We need to establish a robust mental model for users [103], enabling them to instantaneously link the Cymatics Cup with the act of consuming beverages upon seeing it. Furthermore, as we are committed to aligning the visual presentation of beverages with previous tests, the source of vibration is still chosen to be a speaker rather than a vibration motor. A speaker enables precise rendering of shapes corresponding to different frequencies (e.g., adjustable shapes at 27.5Hz or the complete display of all patterns of Cymatics Seasoning), aligning with the results of the user study. Insights from open comments in the user study also revealed that the positive impact of vibration in the oral cavity is limited. Therefore, we need to balance shaping the beverage with vibration while minimizing the potential negative effects on the user experience. Additionally, comments from participants highlighted concerns about beverage splatter, which will also be considered a focal point in the design of the Cymatics Cup.

The design ethos of the Cymatics Cup synthesizes traditional aesthetics with modern mechanical engineering, drawing inspiration from the quintessential Japanese sake container known as the "MASU [46]." In Japanese cultural practices, the act of overpouring sake serves as a tangible expression of hospitality [6]. The Masu functions as a receptacle positioned under the sake cup to capture the overflowing liquid. The integration of the Masu's stylistic language into our prototype is based on both mechanical optimization and enhanced user interaction [31].

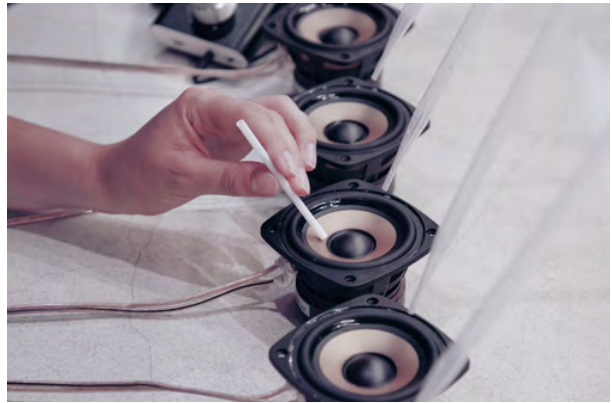

Figure 9: Z-Modena MK2 full-range speaker as a drinking container

*5.2.1 Mechanical Design: An Integrated Engineering Perspective.* Regarding the mechanical architecture, the Masu-inspired form effectively conceals essential components, including the speaker, wiring, driving mechanism, and shock-absorbing material, within the structural base. This design strategy enhances component integrity by safeguarding delicate elements like the speaker and electrical wiring from potential risks posed by direct liquid contact (Fig. 10, depicting the 1/2 cut structure of the design). This encapsulation serves a dual purpose: aesthetic concealment and structural protection, resulting in a seamless fusion of form and function [64].

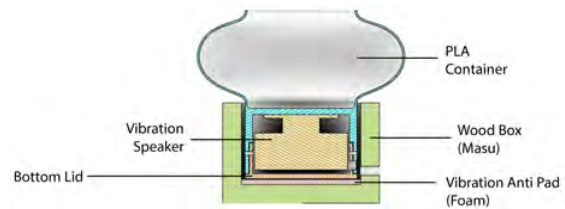

Figure 10: 1/2 cut structure of Cymatics Cup

*5.2.2 User Experience: An Anthropological Insight into Design.* From comments in user study 2 where participants couldn't immediately perceive the speaker as a drinking vessel (P16 and P19), we recognized the importance of incorporating a user-centered interaction and design psychology perspective. Given that the project's objective requires users to taste the liquid, leveraging the familiarity associated with the Masu format establishes an enhanced user interface [47]. This contrasts sharply with the disorienting effect of directly exposing electronic components, such as a speaker, to end-users (Fig. 11, illustrating the comparison between an exposed speaker and our design). Therefore, our Masu-inspired design not only serves as a novel aesthetic but also functions as a cognitive bridge, enhancing the interaction between humans and technology.

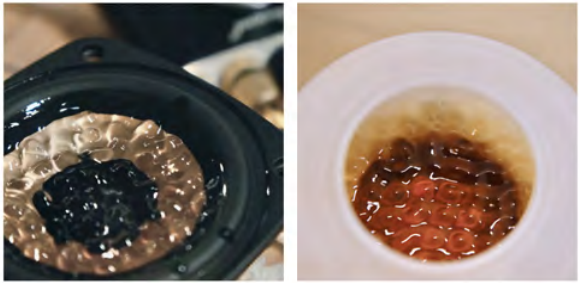

**Figure 11: The comparison between an exposed speaker and Cymatics Cup**

*5.2.3 Material and Component Selection: Acoustic Engineering Analysis.* Material and component selection plays a pivotal role in the product's performance. We conducted a comprehensive comparative analysis of multiple speaker types before finalizing the vibration speaker [89]. Before proceeding with the design of the device, we used two different specifications of conical loudspeakers (Z-Modena MK2 and PLS-P830985 Full Range Speaker), and placed the beverage directly onto the diaphragm for users to conduct early-stage testing. This method helped us verify that the specifications of the loudspeaker had no impact on the generation of liquid surface patterns, while also allowing us to perform the earliest tests at a faster pace.

Subsequently, we chose the vibration speaker. This variety is particularly adept at converting audio signals into tangible surface vibrations, thereby facilitating the creation of intricate water surface patterns. Due to the requirements for the sealing integrity of the loudspeaker body, we ultimately used the vibration loudspeaker. However, according to the feedback from participants P1, P2, P3, P4, P6, P7, and P8 in user study 1, vibrations had a less favorable influence on taste perception when compared to visual effects. Therefore, it is essential to carefully consider how to effectively mitigate the potential impact of vibrations on taste perception while maintaining liquid vibration to create patterns. To mitigate undesirable vibrational patterns—particularly at the base and the sides—we incorporated sound-absorbing materials and ensured optimal spatial positioning to regulate vibrational diffusion [106].

*5.2.4 Container Shape Design: Hydrodynamic Considerations.* While determining the geometry of the upper container, we conducted multiple shape experiments to identify the most effective design (Fig. 12 shows the process of shape design). In phase (1), concerning feedback from participants P4 and P8 regarding beverage splashing issues in user study 2, we tested designs with both open and semi-enclosed shapes and found that the semi-enclosed container shape was more effective in preventing liquid splashing. Therefore, we continued to use this design language in subsequent designs. In phase (2), we aimed to test different specifications of loudspeakers and imposed size limitations, adding a flange at the bottom of the box to secure the loudspeaker and a window for connecting cables to the outside.

In phase (3), after replacing the sound unit with a vibration loudspeaker, we again adjusted the dimensions according to the chosen loudspeaker and designed a structure to adhere the loudspeaker to the container, as opposed to the previous design where it was fixed to the base. This resulted in a better liquid surface pattern effect. After switching to a better-sealed loudspeaker, we once again adjusted the size of the container to the state shown in step (4) and conducted tests.

A new issue arose at this point: this vibration loudspeaker has good vibration conduction properties, causing the sample shown in (4) to move uncontrollably on the table during testing, including the box part. Therefore, we optimized the design to what is shown in step (5), adding a sealed cavity beneath the container and fixing the loudspeaker inside it while also including a vibration-damping sponge between the bottom wooden box and the container to reduce vibration transmission to the table.

This version of the model has achieved the functionality needed for our current research phase, and the establishment of this solution serves a dual purpose. Firstly, it minimizes liquid splashing, thereby enhancing user safety. Secondly, it maximizes the container's surface area, facilitating efficient propagation of sound waves. This design feature is particularly relevant to our research, which aims to investigate the relationship between acoustic vibrations and the dynamics of liquid surfaces [109].

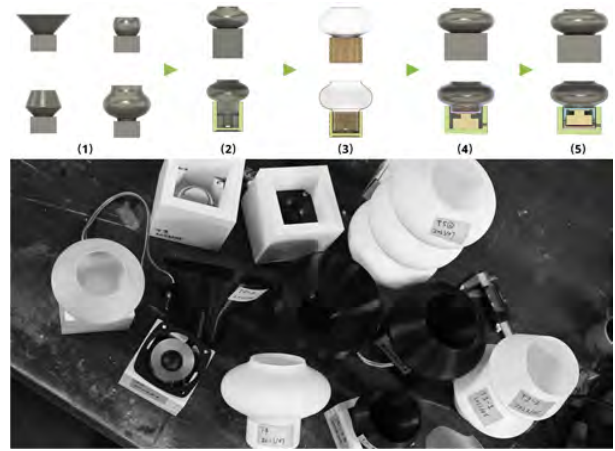

**Figure 12: The process of design**

## 6 LIMITATION AND DESIGN IMPLICATIONS
### 6.1 Limitations
In our user study 1, during the first phase where participants were instructed to observe changes in the shape of solutions, we did not precisely control whether the solution levels in the speaker setup were consistently maintained at the same height due to variations in the amount of liquid consumed by individual participants. Consequently, the sizes of the patterns also varied to some extent. Furthermore, despite listening to white noise as background sound to minimize the impact of external audio, some participants, particularly those sensitive to sound, reported that they could still faintly hear some sound from the speaker. Finally, while many studies use lemon juice and coffee as tests for sourness and bitterness [11, 12, 16], these substances still carry the inherent risk of confounding variables, such as olfactory factors. Although using the same ingredients in the initial mode and subsequent experiments helps mitigate this risk, the presence of confounding factors has its potential. In future experiments, we intend to actively address these concerns, such as by incorporating citric acid and caffeine [65, 100].

In our user study 2, the bitterness test was conducted using Chinese holly leaf, a highly bitter Chinese medicinal herb. This made it challenging for participants to discern taste differences between the baseline, Cymatics Seasoning, and cymatics pop music. From this, it is evident that in future related experiments, it is crucial to select ingredients with moderate flavour intensity. This ensures that participants have greater flexibility to make comparisons across different variables. Moreover, since all taste stimuli used in user study 2 were commercialized products, the complex composition may result in the overall taste presentation differing from the main taste items intended for testing. Therefore, future research should select ingredients with simpler compositions for testing to enhance accuracy. In addition to the choice of ingredient, some participants mentioned that during the Cymatics Seasoning experience, interruptions occurred when there were gaps in the rhythm between frequencies, suggesting the need for a more stable output. A stable output is expected to enhance the seasoning's effectiveness. Future work will focus on adjusting the Cymatics Seasoning to ensure a stable output when applied to the Cymatics Cup.

## 6.2 Design implications

*6.2.1 Difference between Dynamic and Static Shapes.* While we have obtained a series of research results regarding the dynamic shape changes generated by cymatics, we have yet to compare these with static shapes. In future research efforts, we aim to explore the differences between dynamic and static shapes by immobilizing cymatics-generated dynamic shapes, such as freezing certain dynamic particles, to investigate how dynamic and static shapes impact taste perception under identical forms. There are currently two approaches we have in mind for this exploration. One involves patterning cymatics patterns and conducting user studies on taste associations. The alternative approach builds upon the technology of 3D-printed lemon juice gels [105] and involves 3D modeling of cymatics shapes, transforming them into solid forms for in-depth taste-related research, thereby delving deeper into the potential of cymatics.

*6.2.2 Anti-Vibration.* Findings from user study 1 revealed that the introduction of vibrations through various frequencies not only failed to enhance taste perception but, in some cases, interfered with and diminished flavour perception. Despite our efforts to integrate anti-vibration components into the existing prototype of the Cymatics Cup, there persists a residual level of vibration that cannot be entirely eliminated. Consequently, the use of a straw for beverage consumption continues to present a challenge. This challenge arises from the contact between the straw and the surface of the Cymatics Cup, resulting in the transmission of vibrations to the user. For future iterations, it is imperative to optimize the design and functionality of the Cymatics Cup toolkit. We need to conduct further research on anti-vibration materials [79] or refine the structural design between the cup and the straw for optimization.

*6.2.3 Optimization of the Cymatics Cup.* The current prototype of the Cymatics Cup is still in its initial stage. The present Cymatics Cup requires external wiring and an amplifier to control the generation of different Cymatics Seasonings. In the next version of the Cymatics Cup, we plan to incorporate wireless technology, similar to Virtual Lemonade [68] and Augmented Glass [49], in order to eliminate the inconvenience caused by external wiring. Additionally, we have observed that the speaker in this prototype gradually heats up after prolonged use, which is a situation we wish to avoid. Based on safety considerations and to prevent any impact on the temperature of the beverage, in the next version of the Cymatics Cup, we will make efforts to resolve the issue of overheating.

*6.2.4 Cymatics Seasoning Enhancement.* The outcomes of user study 2 brought to light an issue characterized by interruptions in the rhythmic pattern of frequencies, suggesting the need for optimization towards a more stable output rhythm in Cymatics Seasoning. Furthermore, it is worth noting that there are additional dimensions for exploration regarding Cymatics Seasoning. Beyond rhythm, future investigations may delve into the effects of varying tempo, meter, dynamics (volume), vibrato, staccato, and tonality on taste perception. This exploration also extends to potential applications in conjunction with the Cymatics Cup. However, it's important to exercise caution when exploring dynamics variations, as excessive dynamics can result in liquid splattering, potentially leading to user discomfort and a suboptimal user experience. Therefore, meticulous consideration is required when examining dynamic alterations.

*6.2.5 Personalization.* During user study 2, when participants were asked about their expectations regarding the development of the Cymatics Cup's functionality, 19 out of 22 participants expressed their desire to see the Cymatics Cup equipped with the capability to adjust Cymatics Seasonings in the future independently. Therefore, our future endeavors will involve the development of a personalized Cymatics Seasoning system, giving users the capacity to create their own seasonings, similar to a DJ-like bartender. Based on the feedback from user study 1, three participants from different countries mentioned that flavour intensity may be influenced by various cultural factors and personal eating habits. It's worth noting that user study 1 included participants from seven different countries across Asia, Europe, and the Middle East. Therefore, it highlights the importance of personalization and the "self prioritization effect", which can enhance the taste of our food and drinks [76]. This innovative approach aims to enhance the overall enjoyment of the drinking experience. With the comprehensive toolkit offered by the Cymatics Cup, we envision the establishment of a novel drinking ritual.

*6.2.6 Cymatics Recipes.* To establish a comprehensive ecosystem for Cymatics Seasoning, we recognize the importance of collaborating with chefs and nutritionists to create an exhaustive array of recipes. Collaboration with chefs or bartenders in crafting innovative recipes utilizing the Cymatics Cup has the potential to revolutionize the public's drinking experience. Additionally, partnering with nutritionists can significantly enhance dining experiences, particularly for post-operative patients [67] and elderly individuals without teeth [25], who may face challenges in consuming regular meals in a hospital setting. By developing specialized nutritional recipes tailored to the unique needs of these individuals, we can not only stimulate their appetite and enhance their overall dining experience but also address their health concerns. This approach aims to create a friendly and well-being dining environment for hospitalized patients, promoting their physical and emotional well-being.

## 7 CONCLUSION

In this paper, we have presented evidence that cymatics affect our taste perception and overall drinking experience. Specifically, we have demonstrated that the primary impact of cymatics stems from the visual domain, while the tactile effects have a comparatively limited impact on the drinking experience. Furthermore, our exploration of Cymatics Seasoning has revealed its potential to significantly enhance the enjoyment and appetite of users during the drinking experience, aligning with our vision for improved well-being in dining. The development of the Cymatics Cup and the Cymatics Seasoning represents a novel approach to drinking, offering a glimpse into the future of dining rituals, particularly within the field of HCI and the emerging realm of Augmented Eating in the HFI community.

As we look ahead, our future endeavours will focus on refining the toolkit, establishing a robust Cymatics Seasoning system, expanding our user base, and conducting real-world validation studies further to explore the practical applications and benefits of Cymatics Seasoning. Our ultimate aim is to revolutionize the way people experience dining, promoting not only satisfaction but also holistic well-being through the harmonious interplay of sensory elements in the act of eating and drinking.

## ACKNOWLEDGMENTS

The authors wish to extend their sincere appreciation to Danny Hynds for his invaluable contributions to the musical composition in user study 2. Additionally, this research received support from the Japan Science and Technology Agency "Support for Pioneering Research Initiated by the Next Generation (SPRING)" and the 61st Leave a Nest grant "Yoshinoya Award". The assistance provided by these entities was pivotal to the completion of this work.